\documentclass[sigconf]{acmart}
\usepackage[]{xcolor}
\usepackage[normalem]{ulem}
\AtBeginDocument{%
  \providecommand\BibTeX{{%
    \normalfont B\kern-0.5em{\scshape i\kern-0.25em b}\kern-0.8em\TeX}}}
\usepackage{algorithm,algorithmic}
\setcopyright{acmcopyright}
\copyrightyear{2022}
\acmYear{2022}
\acmDOI{10.1145/1122445.1122456}

\acmConference[Phoenix '22]{Phoenix '22: the 15th ACM International WSDM Conference}{Feb 21--25, 2022}{Phoenix, AZ}
\acmBooktitle{Phoenix '22: the 15th ACM International WSDM Conference,
  Feb 21--25, 2022, Phoenix, AZ}
\acmPrice{15.00}
\acmISBN{978-1-4503-XXXX-X/18/06}

\begin{document}

\title{Query Expansion and Entity Weighting for Query Reformulation Retrieval in Voice Assistant Systems}

\author{Zhongkai Sun}
\email{zhongkas@amazon.com}
\affiliation{%
  \institution{Amazon Alexa AI}
  \city{Seattle}
  \state{WA}
  \country{USA}
}

\author{Sixing Lu}
\email{cynthilu@amazon.com}
\affiliation{%
  \institution{Amazon Alexa AI}
  \city{Seattle}
  \state{WA}
  \country{USA}}

\author{Chengyuan Ma}
\email{mchengyu@amazon.com}
\affiliation{%
  \institution{Amazon Alexa AI}
  \city{Seattle}
  \state{WA}
  \country{USA}}

\author{Xiaohu Liu}
\email{derecliu@amazon.com}
\affiliation{%
  \institution{Amazon Alexa AI}
  \city{Seattle}
  \state{WA}
  \country{USA}}

\author{Chenlei Guo}
\email{guochenl@amazon.com}
\affiliation{%
  \institution{Amazon Alexa AI}
  \city{Seattle}
  \state{WA}
  \country{USA}}




\begin{abstract}
Voice assistants such as Alexa, Siri, and Google Assistant have become increasingly popular worldwide. However, linguistic variations, variability of speech patterns, ambient acoustic conditions, and other such factors are often correlated with the assistants misinterpreting the user's query. In order to provide better customer experience, retrieval based query reformulation (QR) systems are widely used to reformulate those misinterpreted user queries. Current QR systems typically focus on neural retrieval model training or direct entities retrieval for the reformulating. However, these methods rarely focus on query expansion and entity weighting simultaneously, which may limit the scope and accuracy of the query reformulation retrieval. In this work, we propose a novel Query Expansion and Entity Weighting method (QEEW), which leverages the relationships between entities in the entity catalog (consisting of users' queries, assistant's responses, and corresponding entities), to enhance the query reformulation performance. Experiments on Alexa annotated data demonstrate that QEEW improves all top precision metrics, particularly 6\% improvement in top10 precision, compared with baselines not using query expansion and weighting; and more than 5\% improvement in top10 precision compared with other baselines using query expansion and weighting.

\end{abstract}




\maketitle

\section{Introduction}
Virtual voice assistants such as Alexa, Siri, and Google Assistant have become daily necessities for worldwide users. There are two main components in such systems, namely Automatic Speech Recognition (ASR) and Natural Language Understanding (NLU) to extract the semantic information from an input voice query. However, misinterpretations may occur owing to factors such as semantic/speech pattern variants and ambient acoustic conditions, resulting in incorrect responses and unsatisfactory user experience. Therefore, it has become a vital yet challenging task to enable voice assistants to correctly understand those misinterpreted user queries.

Many works have been devoted to query reformulating (QR) to correct misinterpreted user queries \cite{wang2021contextual, fan2021search, buck2017ask, cao2008context, xu2017quary, he2016learning}. Among QR systems, the retrieval and ranking based approaches are efficient and widely-used in the industry \cite{fan2021search, wang2021contextual, cho2021personalized}. These works focus more on retrieval models without additional entity information. However, in voice assistant systems where task-oriented queries are usually short and ambiguous, the query intention may not be fully interpreted and this limitation negatively impact the performance of QR system whose relevance strongly relies on the query. For instance, a user's input query ("play long distance love by Sheena Easton") contains a correct singer name ("Sheena Easton") but also a lyric ("long distance love") instead of the correct song name ("telefone"). To reformulate this query so that virtual voice assistants can find the target song to play, the QR system needs to utilize the existing entities to obtain the correct song name "telefone"; besides, the model also needs to make a decision between two candidates ("play long distance love by little feat" and "play telefone by sheena easton") as they both contain key words ("long distance love" or "sheena easton") in the original query.


Query expansion is an effective approach to enrich the original oral query. Some researches consider the correlations between a query and its expansion candidates to select the most relevant candidate from the candidate pool: \cite{cui2002probabilistic} proposes a query expansion method based on probabilistic correlations between query logs; \cite{imani2019deep} uses a DNN model to directly predict the scores for query expansion terms. Both approaches make query more complete but do not consider the weighting between original query and expanded ones. Other works have studied how to directly generate the expansions: \cite{gao2012towards} proposes a concept-based translation method to generate the expanded query; \cite{mustar2020using} uses BERT \cite{devlin2018bert} and BART \cite{lewis2019bart} with a next query prediction task to do query suggestion; docT5query \cite{nogueira2019doc2query} implements a powerful T5 model \cite{raffel2019exploring} to predict the relevant query for each document. The generative approaches are able to provide large number of variants, but it may also bring noise and lead to incorrect reformulations.

Entity (term/phrase) weighting is also a common approach to emphasize the key words in the query and improve the retrieval performance. To predict the weight of each entity, conventional methods use the corpus statistics based methods like TF-IDF, BM25, and SRNM \cite{jones1972statistical, zamani2018neural} to compute the scores of entities. Recently, neural network based methods have been widely studied: \cite{roitman2019query} uses a supervised query performance prediction method to learn query term weighting; \cite{zhang2017mike} utilizes a random-walk model to learn latent representations for candidates to capture the relative importance between entities;
\cite{bai2020sparterm} uses pre-trained language models to learn sparse entity representations in order to better predict the importance of each entity. However, these works haven't utilized the expansion and term weighting simultaneously. DeepImapct \cite{mallia2021learning} first expands the documents using docT5query \cite{nogueira2019doc2query}, and then uses a contrastive loss to predict the weight of each term in the query and document. However, DeepImapct focuses on generating the expansions for long documents (e.g. a paragraph), but it has not been expanded to short query scenarios.


In this paper, we integrate query expansion and entity weighting together to propose a novel method called \textbf{Q}uery \textbf{E}xpansion and \textbf{E}ntity \textbf{W}eights prediction (QEEW) for query reformulation. Specifically, QEEW first establishes an entity Expansion Knowledge Base (EEKB) by categorizing and linking entities in an entity catalog (which consists of user queries, assistant's responses and corresponding entities) to expand queries. An entity weighting prediction model is then trained using annotated query reformulation pairs with expansions obtained from the EEKB. By integrating query entity weighting with query expansion, QEEW can retrieve reformulations more relevant to the user's intention. Our evaluation on annotated Alexa data shows that the proposed QEEW improves  2.6\%, 5.3\%, 1.1\% on precision@50, precision@10, and precision@1 compared with best scores obtained from using other methods.


\section{Methodology}
\begin{figure}[h]
  \centering
  \includegraphics[width=\linewidth]{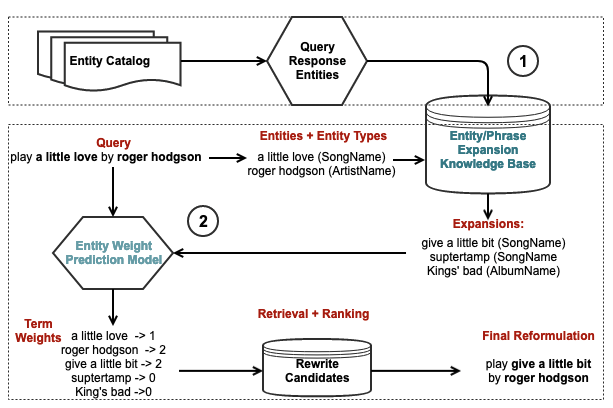}
    \caption{The overview of the workflow:
    1) entity catalog data is first used to build the Entity Expansion Knowledge Base; 2) an entity weight prediction model is trained with annotated Alexa <query, reformulation> pairs.}
  \label{fig:pipeline}
\end{figure}

This section describes the details of the proposed QEEW method. The training data are pairs of annotated misinterpreted queries and their corresponding reformulations, and the EEKB is built from a pre-established user-satisfied entity catalog. To determine if a user's query is successfully processed by the voice assistant, an automatic friction estimation system \cite{gupta2021robertaiq} is applied, and only the non-frictional queries and their corresponding assistant's responses in the catalog will be used to build the EEKB. As shown in Figure \ref{fig:pipeline}, our proposed QEEW model mainly consists of two components: 
\begin{itemize}
    \item \textbf{Entity Expansion Knowledge Base} 
    An EEKB which contains connections between different entities is built by categorizing entities based on their occurrences in queries  and responses, both of which are obtained form the user satisfied entity catalog.
    Once this knowledge base is built, it will be used to provide expansion candidates in both training and test steps. Details of the building EEKB and how to expand query with it will be introduced in section 2.1.
    \item \textbf{Entity Weight Prediction Model} After expanding the query from the EEKB, an entity weight prediction model will be trained to predict the weighting scores of both entities in the original query and those expansions obtained from the EEKB. Section 2.2 introduces the details of this entity weight prediction model.
\end{itemize}

The expanded query with the predicted entity weights can then be used in the downstream retrieval/ranking QR system to obtain the final reformulation.

\begin{figure}[h]
  \centering
  \includegraphics[width=\linewidth]{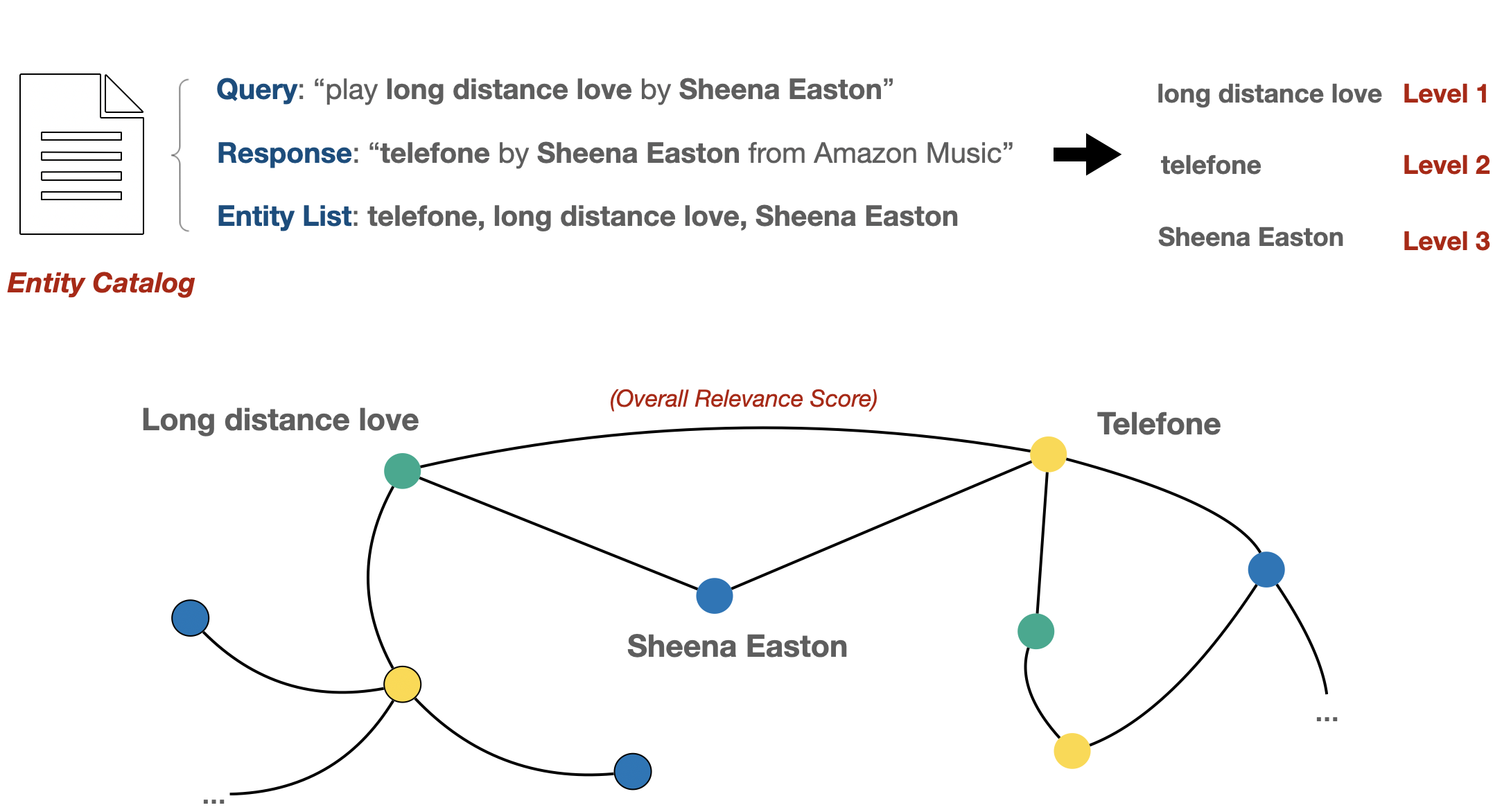} \caption{ Illustration of the building process of the entity connection knowledge base. First, each entity in one entity catalog entry will be categorized based on its occurrence in this entry's query and response. Second, every two entities in the same entry will be connected with each other. Third, for each edge, the "overall relevance score" is calculated based on these two entities' respective scores in all times they appear together.}
  
  \label{fig:mapdict}
\end{figure}


\begin{figure}[h]
  \centering
  \includegraphics[width=\linewidth]{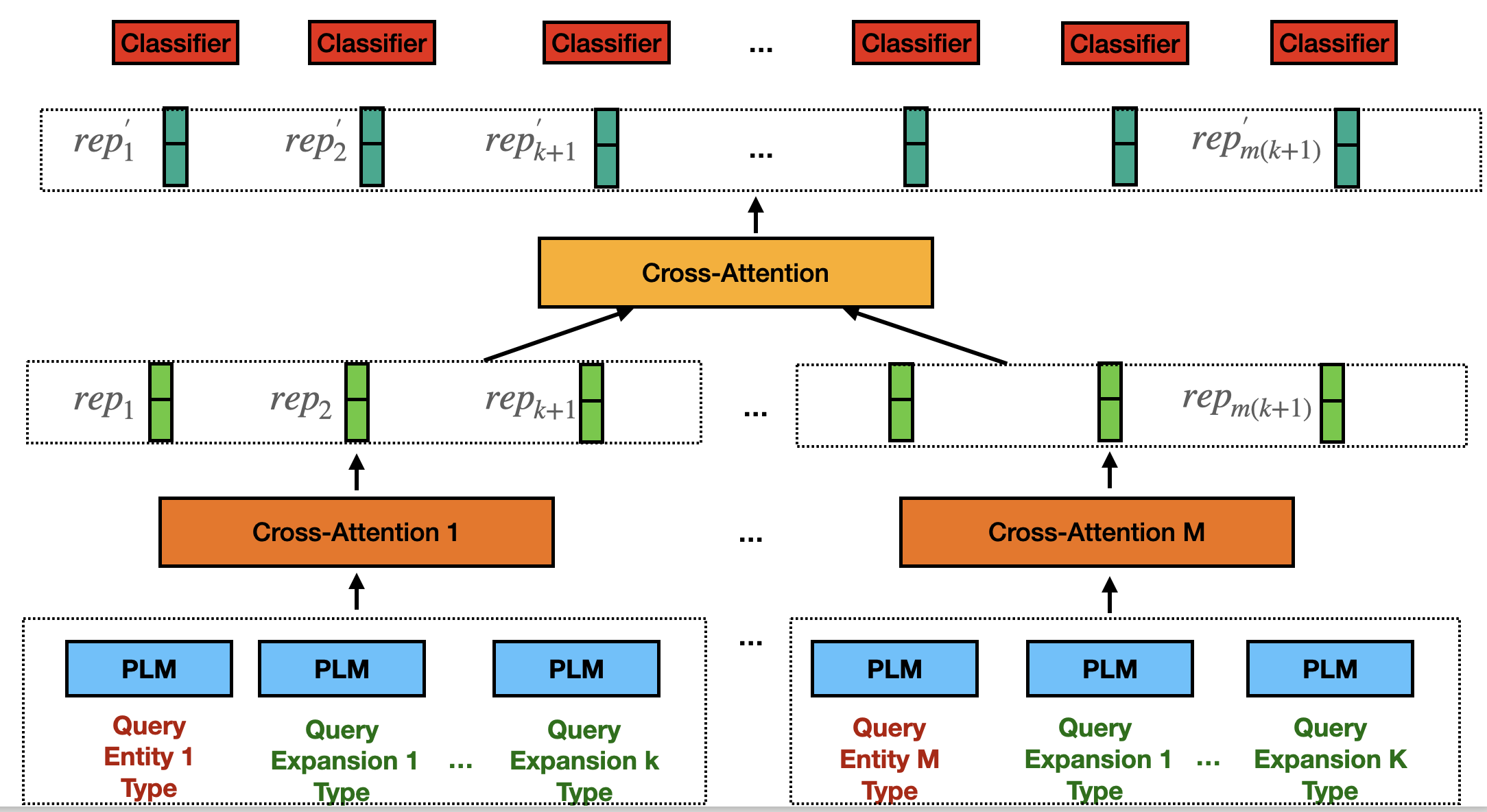}
  \caption{The illustration of the entity weight prediction model. A hierarchical structure is applied to predict the relative weights for all entities}
  \label{fig:model}
\end{figure}

\subsection{Entity Expansion Knowledge Base}


The EEKB is built using a comprehensive high-quality entity catalog consisting of user's queries, user-satisfied assistant's responses, and every query-response pair's unique entity list. This entity catalog is able to provide corrections for those inaccurate entities; and the voice assistant's responses can also help to indicate entities' relative importance. 
By connecting each entity with others and categorizing each entity according to its existence in query and response, the EEKB can be established with the relationships between different types of entities.
Therefore, a query can be expanded by finding the most relevant expansions in this EEKB for every entity in the query.

\begin{algorithm}
\begin{algorithmic}
\renewcommand{\algorithmicrequire}{\textbf{Input:}}
\REQUIRE \textbf{ENTITY CATALOG} $\mathbb{C}=\{(C_{i}\}$, $1 \leq i \leq K$, \\
where $C_{i} = \{Query_{i}, Response_{i}, E_{i} \}$, \\
where $E_{i} \gets unique entities in \{Query_{i}, Response_{i} \}$  
\\
\STATE KG $\leftarrow \{\}$ \\
\FOR{$i \in \{1,\dots,K\}$}
      \FOR{$e_{j}$ in $E_{i}$}
        \IF{$e_{j}$ in $Query_{i}, Response_{i}$} 
            \STATE $Level_{j} \gets 3$
        \ELSIF{$e_{j}$ in $Response_{i}$ }
            \STATE $Level_{j} \gets 2$
        \ELSE
            \STATE $Level_{j} \gets 1$       
        \ENDIF
      \ENDFOR
      
      \FOR{$e_{m},e_{n}$ in $E_{i}, m \neq n$}
            \STATE 
            KG[$e_{m}$][$e_{n}$] += $level_{m} \times level_{n}$ 
            \STATE
            KG[$e_{n}$][$e_{m}$] += $level_{n} \times level_{m}$
      \ENDFOR
\ENDFOR
\end{algorithmic}
\caption{Building the entity expansion knowledge base}
\label{algortihm:kb}
\end{algorithm}
Figure \ref{fig:mapdict} demonstrates the building process of the EEKB. Nodes in the EEKB are entities from the entity catalog, and each node is connected with others when both nodes appear in the same entry in the entity catalog. The score on the edge indicates the overall relevance level of two nodes. For an entity in one entry, its current relevance level is based on its occurrence in query and response. Level 1 means that this entity only exists in the query; level 2 represents that it only exists in the response; and level 3 indicates that it exists in both query and response. The motivation is that the higher the relevance level, the more relevant the corresponding entity with the query is. In the example in Figure \ref{fig:mapdict}, the entity "Sheena Easton" has a relevance level 3 because it exists in both query and response. Every two entities' relevance levels are also recorded. For instance, in Figure \ref{fig:mapdict}, $"telefone"$ and $"Sheena Easton"$ will have a relevance level pair $[level 2, level 3]$. The overall relevance level score between two connected nodes is defined as the sum of the product of the their respective scores each time they appear together. The detailed algorithm is presented in Algorithm \ref{algortihm:kb}.

Once EEKB is established, it can be used to expand a specific query. For each entity in the original target query, its K most relevant nodes (ranked according to the overall relevance score) is obtained from the knowledge base. However, since typically only limited expansions are useful for reformulation retrieval, the weighting model trained in 2.2 can help to polish the expansion. For example, some of the expanded phases may be assigned weight=0 and will be removed from the expansion.


\subsection{Entity Weights Prediction Model}

The weight prediction model is trained for entities in the original query as well as expansions obtained from the EEKB in section 2.1. The training data is the labelled Alexa query reformulation pairs, and the weight to be predicted for each entity is a specific weighting score level based on the reformulation. 

As shown in the workflow in Figure \ref{fig:pipeline}, the task is defined as predicting the weighting score level of all entities (both original and expanded) for a given query. Entities exist in the reformulation will have higher weighting score levels while those not in the reformulation will have lower score levels. The weighting score level of each entity is assigned according to the existence in the labelled Alexa query and reformulation pairs: 1) if this entity exist neither in the original defective query nor in the reformulation, the score level is labelled as 0; 2) the score level of entities in original query will be labeled as 1; 3) the score level of entities in reformulation (either in request or not) will be labeled as 2, which indicates the most important weight.

The weight training task is defined as: give a query $Q$ with $m$ entities $E_{1}, E_{2},..., E_{m}$ and each entity $E_{i}$ has $k$ expansions $E_{i}^{1}, E_{i}^{2},..., E_{i}^{k}$, the target is to predict the importance score $l_{p}$ for every entity $E_{p}$ (both original and expanded) where $1 \leq p \leq m(k+1)$. 
In this work,  we train a hierarchical model to predict each entity's weight. Figure \ref{fig:model} demonstrates the structure of the hierarchical model, which can better learning the relationships between each entity and its expansions instead of predicting all of the weights in one step.

As shown in Figure \ref{fig:model}, a pretrained language model (PLM) is used to encode each entity $E_p$. Specifically, each $E_p$ is concatenated with the original query and its entity type (e.g., "SongName", "AuthorName") to form the input text "Query [SEP] Entity [SEP] Type". This input text will be encoded to an embedding by the PLM. After that, $M$ first-layer cross-attentions modules are applied to $M$ raw entity and their expansions. So that each first-layer cross-attention module can learn the relationships between each raw entity and its expansions. At last, each entity's representation $rep_{p}$ (where $1 \leq p \leq m(k+1) $) from all first cross-attention modules are jointly input to the second cross attention module. Therefore, the interactions between all entities can be learned and the outputs $rep_{p}^{'}$ (where$1 \leq p \leq m(k+1) $) will then be input to the final classification layer to predict the importance weight logits $g_{p}$. Cross-entropy is used to calculate the loss between each entity's predicted logits $g_p$ and its true label $l_p$, the final loss is defined as the average of these losses:
\begin{equation}
    \mathcal{L}_{final} = \frac{1}{m(k+1)} \sum_{p=1}^{m(k+1)} \text{CrossEntropy} (l_p, g_p)
     \label{eq:loss}
\end{equation}


\section{Experiments and Evaluation}
We extract 2000 hours of successfully processed Alexa data from an aggregated and anonymized datalog (which has no user specific information), and use it to build the EEKB and the retrieval index. An annotated query-reformulation dataset is split into 60K/15k/27k, to be used for training, evaluation, and testing, respectively.

The Robert-Base\cite{liu2019roberta} model is utilized as the encoder and a multi-head attention module is used as the cross-attention layer in Figure \ref{fig:model}. Hyper-parameters of the model can be found in Appendix A. P@K (Precision of top-K retrieved candidates) is used as the evaluation metrics, and four baselines are used for the  comparison. \\
1) The ElasticSearch\footnote{Version 7.10 is used in this work}, which is a popular string match retrieval method based on the inverted indexing and BM25 \cite{jones1972statistical}; \\
2) The SimCSE \cite{gao2021simcse}, which is a neural network based model that leverages the contrastive learning to build a representation encoder; \\
3) DocT5query \cite{nogueira2019doc2query}, which learns to expand the document by predicting its relevant query using the T5 model. The expanded document can be used in a specific retrieval system, e.g. Elasticsearch.  \\
4) DeepImpact \cite{mallia2021learning}, which performs both document expansion and semantic importance estimation using DocT5Query and a contrastive learning based term prediction. 
\begin{table}
  \label{tab:freq}
  \begin{tabular}{|c|c|c|l|}
    \toprule
    Method & P@50 & P@10 & P@1 \\
    \midrule
    DeepImapct &63.8\% & 32.4\% & 16.3\% \\
    Elasticsearch &85.4\% & 67.6\% &20.6\% \\
    docT5query on Elasticsearch &85.6\% & 68.3\% & 20.5\% \\
    QEEW on Elasticsearch  & \textbf{88.2\%} & \textbf{73.6\%} & 25.1\% \\
    SimCSE & 72.7\% & 64.0\% & 33.7\% \\
    QEEW on SimCSE & 75.9\% & 64.1\% & \textbf{34.8}\% \\
  \bottomrule
\end{tabular}
\caption{Main result table. QEEW on Elasticsearch is able to achieve significantly better performance on P@50 and P@10; QEEW on SimCSE achieves the best performane on P@1.}
\label{tab:mainresult}
\end{table}
\begin{table}
  \label{tab:freq}
  \begin{tabular}{|c|c|c|l|}
    \toprule
    Method & P@50 & P@10 & P@1 \\
    \midrule
     Elasticsearch &85.4\% & 67.6\% &20.6\% \\
    Elasticsearch with expansion &86.3\% & 68.5\%& 21.1\% \\
    Elasticsearch with weight &87.4\% & 73.1\% &25.0\% \\
    QEEW on Elasticsearch & \textbf{88.2\%} & \textbf{73.6\%} & \textbf{25.1\%} \\
  \bottomrule
\end{tabular}
\caption{The ablation study on Elasticsearch demonstrates the usefulness of both expansion and weight prediction.}
\label{tab:ablaton}
\end{table}

To evaluate the effectiveness of QEEW, we applied it to 1) string-match based Elasticsearch and 2) the neural embedding based SimCSE. On Elasticsearch, we first build the reformulation index from reformulation candidates set, and append each query with its expanded entities to form a new query. The predicted weights are applied to the retrieval score function so that the higher weighted entities contribute to higher scores for ranking; 
On SimCSE, the retrieval of ranking reformulations is based on the similarity score between query and reformulation strings, and the relevance score is the Euclidean Distance between the embeddings of query and reformulations. To apply the QEEW, we also adjust the similarity scores of retrieval reformulations based on the weights so that the reformulations that contain higher weighted entities will be with higher ranking scores.

Table \ref{tab:mainresult} shows that the proposed QEEW method can achieve the highest P@50 and P@1 when applied to Elasticsearch, and achieves the best P@1 when applied to the SimCSE. Specifically, QEEW can improve P@50, P@10, and P@1 by 2.8\%, 6\%, and 4.5\% respectively in Elasticsearch and by 3.2\%, 0.1\%, and 1.1\% respectively in SimCSE. Notice that the SimCSE achieves higher P@1 but lower P@10/P@50 compared with Elasticsearch, this is because that the SimCSE is trained with a semantic matching objective so that it performs better on precise matching, while Elasticsearch enables a wider search and thus achieves higher P@50 and P@10.
When comparing with (3) DocT5Query and (4) DeepImapct, our proposed method still achieves better performance. Specifically, Elasticsearch on QEEW can improve 2.6\%, 5.3\%, and 4.6\% on P@50, P@10, and P@1 respectively compared to docT5query on Elasticsearch; QEEW on SimCSE can improve 12.2\%, 31.7\%, and 18.5\% on P@50, P@10, and P@1 respectively compared to DeepImapct. The reason why (3) and (4) cannot achieve as good performance as our proposal is that these models are more suitable for longer paragraphs but fail to perfectly expand to the short and ambiguous spoken queries.

Table \ref{tab:ablaton} demonstrates the ablation study result of applying QEEW to Elasticsearch. Elasticsearch with predicted expansions singly is able to gain 0.9\%, 0.9\%, 0.5\% on P@50, P@10, and P@1; while with predicted weights singly can improve 2\%, 5.5\%, 4.4\% on P@50, P@10, and P@1. This result demonstrates weighting can significantly improve the performance compared to query expansion alone.
\section{Conclusions and Future Work}

In this work, we propose a novel method named as QEEW for query expansion and entity weighting predicion. Experiments on annotated Alexa data demonstrates that 1) QEEW can increase retrieval precision when applying to both conventional string-matching retrieval system (e.g., Elasticsearch) and novel neural network retrieval system (e.g., SimCSE); 2) QEEW achieves better performance compared with SOTA retrieval based QR systems (e.g., DocT5Query, DeepImpact).


For the future work, we would like to investigate more effective ways to learn entity extension/ weighting and how to leverage the EEKB and predicted entity weights to improve other tasks such as recommendation and personalization.

\clearpage



\bibliographystyle{ACM-Reference-Format}
\bibliography{sample-base}


\begin{thebibliography}{25}


\ifx \showCODEN    \undefined \def \showCODEN     #1{\unskip}     \fi
\ifx \showDOI      \undefined \def \showDOI       #1{#1}\fi
\ifx \showISBNx    \undefined \def \showISBNx     #1{\unskip}     \fi
\ifx \showISBNxiii \undefined \def \showISBNxiii  #1{\unskip}     \fi
\ifx \showISSN     \undefined \def \showISSN      #1{\unskip}     \fi
\ifx \showLCCN     \undefined \def \showLCCN      #1{\unskip}     \fi
\ifx \shownote     \undefined \def \shownote      #1{#1}          \fi
\ifx \showarticletitle \undefined \def \showarticletitle #1{#1}   \fi
\ifx \showURL      \undefined \def \showURL       {\relax}        \fi
\providecommand\bibfield[2]{#2}
\providecommand\bibinfo[2]{#2}
\providecommand\natexlab[1]{#1}
\providecommand\showeprint[2][]{arXiv:#2}

\bibitem[\protect\citeauthoryear{Bai, Li, Wang, Zhang, Shang, Xu, Wang, Wang,
  and Liu}{Bai et~al\mbox{.}}{2020}]%
        {bai2020sparterm}
\bibfield{author}{\bibinfo{person}{Yang Bai}, \bibinfo{person}{Xiaoguang Li},
  \bibinfo{person}{Gang Wang}, \bibinfo{person}{Chaoliang Zhang},
  \bibinfo{person}{Lifeng Shang}, \bibinfo{person}{Jun Xu},
  \bibinfo{person}{Zhaowei Wang}, \bibinfo{person}{Fangshan Wang}, {and}
  \bibinfo{person}{Qun Liu}.} \bibinfo{year}{2020}\natexlab{}.
\newblock \showarticletitle{SparTerm: Learning Term-based Sparse Representation
  for Fast Text Retrieval}.
\newblock \bibinfo{journal}{\emph{arXiv preprint arXiv:2010.00768}}
  (\bibinfo{year}{2020}).
\newblock


\bibitem[\protect\citeauthoryear{Buck, Bulian, Ciaramita, Gajewski, Gesmundo,
  Houlsby, and Wang}{Buck et~al\mbox{.}}{2017}]%
        {buck2017ask}
\bibfield{author}{\bibinfo{person}{Christian Buck}, \bibinfo{person}{Jannis
  Bulian}, \bibinfo{person}{Massimiliano Ciaramita}, \bibinfo{person}{Wojciech
  Gajewski}, \bibinfo{person}{Andrea Gesmundo}, \bibinfo{person}{Neil Houlsby},
  {and} \bibinfo{person}{Wei Wang}.} \bibinfo{year}{2017}\natexlab{}.
\newblock \showarticletitle{Ask the right questions: Active question
  reformulation with reinforcement learning}.
\newblock \bibinfo{journal}{\emph{arXiv preprint arXiv:1705.07830}}
  (\bibinfo{year}{2017}).
\newblock


\bibitem[\protect\citeauthoryear{Cao, Jiang, Pei, He, Liao, Chen, and Li}{Cao
  et~al\mbox{.}}{2008}]%
        {cao2008context}
\bibfield{author}{\bibinfo{person}{Huanhuan Cao}, \bibinfo{person}{Daxin
  Jiang}, \bibinfo{person}{Jian Pei}, \bibinfo{person}{Qi He},
  \bibinfo{person}{Zhen Liao}, \bibinfo{person}{Enhong Chen}, {and}
  \bibinfo{person}{Hang Li}.} \bibinfo{year}{2008}\natexlab{}.
\newblock \showarticletitle{Context-aware query suggestion by mining
  click-through and session data}. In \bibinfo{booktitle}{\emph{Proceedings of
  the 14th ACM SIGKDD international conference on Knowledge discovery and data
  mining}}. \bibinfo{pages}{875--883}.
\newblock


\bibitem[\protect\citeauthoryear{Cho, Jiang, Hao, Chen, Gupta, Fan, and
  Guo}{Cho et~al\mbox{.}}{2021}]%
        {cho2021personalized}
\bibfield{author}{\bibinfo{person}{Eunah Cho}, \bibinfo{person}{Ziyan Jiang},
  \bibinfo{person}{Jie Hao}, \bibinfo{person}{Zheng Chen},
  \bibinfo{person}{Saurabh Gupta}, \bibinfo{person}{Xing Fan}, {and}
  \bibinfo{person}{Chenlei Guo}.} \bibinfo{year}{2021}\natexlab{}.
\newblock \showarticletitle{Personalized Search-based Query Rewrite System for
  Conversational AI}. In \bibinfo{booktitle}{\emph{Proceedings of the 3rd
  Workshop on Natural Language Processing for Conversational AI}}.
  \bibinfo{pages}{179--188}.
\newblock


\bibitem[\protect\citeauthoryear{Cui, Wen, Nie, and Ma}{Cui
  et~al\mbox{.}}{2002}]%
        {cui2002probabilistic}
\bibfield{author}{\bibinfo{person}{Hang Cui}, \bibinfo{person}{Ji-Rong Wen},
  \bibinfo{person}{Jian-Yun Nie}, {and} \bibinfo{person}{Wei-Ying Ma}.}
  \bibinfo{year}{2002}\natexlab{}.
\newblock \showarticletitle{Probabilistic query expansion using query logs}. In
  \bibinfo{booktitle}{\emph{Proceedings of the 11th international conference on
  World Wide Web}}. \bibinfo{pages}{325--332}.
\newblock


\bibitem[\protect\citeauthoryear{Devlin, Chang, Lee, and Toutanova}{Devlin
  et~al\mbox{.}}{2018}]%
        {devlin2018bert}
\bibfield{author}{\bibinfo{person}{Jacob Devlin}, \bibinfo{person}{Ming-Wei
  Chang}, \bibinfo{person}{Kenton Lee}, {and} \bibinfo{person}{Kristina
  Toutanova}.} \bibinfo{year}{2018}\natexlab{}.
\newblock \showarticletitle{Bert: Pre-training of deep bidirectional
  transformers for language understanding}.
\newblock \bibinfo{journal}{\emph{arXiv preprint arXiv:1810.04805}}
  (\bibinfo{year}{2018}).
\newblock


\bibitem[\protect\citeauthoryear{Fan, Cho, Huang, and Guo}{Fan
  et~al\mbox{.}}{2021}]%
        {fan2021search}
\bibfield{author}{\bibinfo{person}{Xing Fan}, \bibinfo{person}{Eunah Cho},
  \bibinfo{person}{Xiaojiang Huang}, {and} \bibinfo{person}{Chenlei Guo}.}
  \bibinfo{year}{2021}\natexlab{}.
\newblock \showarticletitle{Search based Self-Learning Query Rewrite System in
  Conversational AI}. In \bibinfo{booktitle}{\emph{2nd International Workshop
  on Data-Efficient Machine Learning (De-MaL)}}.
\newblock


\bibitem[\protect\citeauthoryear{Gao and Nie}{Gao and Nie}{2012}]%
        {gao2012towards}
\bibfield{author}{\bibinfo{person}{Jianfeng Gao} {and}
  \bibinfo{person}{Jian-Yun Nie}.} \bibinfo{year}{2012}\natexlab{}.
\newblock \showarticletitle{Towards concept-based translation models using
  search logs for query expansion}. In \bibinfo{booktitle}{\emph{Proceedings of
  the 21st ACM international conference on Information and knowledge
  management}}. \bibinfo{pages}{1--10}.
\newblock


\bibitem[\protect\citeauthoryear{Gao, Yao, and Chen}{Gao et~al\mbox{.}}{2021}]%
        {gao2021simcse}
\bibfield{author}{\bibinfo{person}{Tianyu Gao}, \bibinfo{person}{Xingcheng
  Yao}, {and} \bibinfo{person}{Danqi Chen}.} \bibinfo{year}{2021}\natexlab{}.
\newblock \showarticletitle{SimCSE: Simple Contrastive Learning of Sentence
  Embeddings}.
\newblock \bibinfo{journal}{\emph{arXiv preprint arXiv:2104.08821}}
  (\bibinfo{year}{2021}).
\newblock


\bibitem[\protect\citeauthoryear{Gupta, Fan, Liu, Yao, Ling, Zhou, Pham, and
  Guo}{Gupta et~al\mbox{.}}{2021}]%
        {gupta2021robertaiq}
\bibfield{author}{\bibinfo{person}{Saurabh Gupta}, \bibinfo{person}{Xing Fan},
  \bibinfo{person}{Derek Liu}, \bibinfo{person}{Benjamin Yao},
  \bibinfo{person}{Yuan Ling}, \bibinfo{person}{Kun Zhou},
  \bibinfo{person}{Tuan-Hung Pham}, {and} \bibinfo{person}{Chenlei Guo}.}
  \bibinfo{year}{2021}\natexlab{}.
\newblock \showarticletitle{RoBERTaIQ: An Efficient Framework for Automatic
  Interaction Quality Estimation of Dialogue Systems}. In
  \bibinfo{booktitle}{\emph{2nd International Workshop on Data-Efficient
  Machine Learning (DeMaL)}}.
\newblock


\bibitem[\protect\citeauthoryear{He, Tang, Ouyang, Kang, Yin, and Chang}{He
  et~al\mbox{.}}{2016}]%
        {he2016learning}
\bibfield{author}{\bibinfo{person}{Yunlong He}, \bibinfo{person}{Jiliang Tang},
  \bibinfo{person}{Hua Ouyang}, \bibinfo{person}{Changsung Kang},
  \bibinfo{person}{Dawei Yin}, {and} \bibinfo{person}{Yi Chang}.}
  \bibinfo{year}{2016}\natexlab{}.
\newblock \showarticletitle{Learning to rewrite queries}. In
  \bibinfo{booktitle}{\emph{Proceedings of the 25th ACM International on
  Conference on Information and Knowledge Management}}.
  \bibinfo{pages}{1443--1452}.
\newblock


\bibitem[\protect\citeauthoryear{Imani, Vakili, Montazer, and Shakery}{Imani
  et~al\mbox{.}}{2019}]%
        {imani2019deep}
\bibfield{author}{\bibinfo{person}{Ayyoob Imani}, \bibinfo{person}{Amir
  Vakili}, \bibinfo{person}{Ali Montazer}, {and} \bibinfo{person}{Azadeh
  Shakery}.} \bibinfo{year}{2019}\natexlab{}.
\newblock \showarticletitle{Deep neural networks for query expansion using word
  embeddings}. In \bibinfo{booktitle}{\emph{European Conference on Information
  Retrieval}}. Springer, \bibinfo{pages}{203--210}.
\newblock


\bibitem[\protect\citeauthoryear{Jones}{Jones}{1972}]%
        {jones1972statistical}
\bibfield{author}{\bibinfo{person}{Karen~Sparck Jones}.}
  \bibinfo{year}{1972}\natexlab{}.
\newblock \showarticletitle{A statistical interpretation of term specificity
  and its application in retrieval}.
\newblock \bibinfo{journal}{\emph{Journal of documentation}}
  (\bibinfo{year}{1972}).
\newblock


\bibitem[\protect\citeauthoryear{Lewis, Liu, Goyal, Ghazvininejad, Mohamed,
  Levy, Stoyanov, and Zettlemoyer}{Lewis et~al\mbox{.}}{2019}]%
        {lewis2019bart}
\bibfield{author}{\bibinfo{person}{Mike Lewis}, \bibinfo{person}{Yinhan Liu},
  \bibinfo{person}{Naman Goyal}, \bibinfo{person}{Marjan Ghazvininejad},
  \bibinfo{person}{Abdelrahman Mohamed}, \bibinfo{person}{Omer Levy},
  \bibinfo{person}{Ves Stoyanov}, {and} \bibinfo{person}{Luke Zettlemoyer}.}
  \bibinfo{year}{2019}\natexlab{}.
\newblock \showarticletitle{Bart: Denoising sequence-to-sequence pre-training
  for natural language generation, translation, and comprehension}.
\newblock \bibinfo{journal}{\emph{arXiv preprint arXiv:1910.13461}}
  (\bibinfo{year}{2019}).
\newblock


\bibitem[\protect\citeauthoryear{Liu, Ott, Goyal, Du, Joshi, Chen, Levy, Lewis,
  Zettlemoyer, and Stoyanov}{Liu et~al\mbox{.}}{2019}]%
        {liu2019roberta}
\bibfield{author}{\bibinfo{person}{Yinhan Liu}, \bibinfo{person}{Myle Ott},
  \bibinfo{person}{Naman Goyal}, \bibinfo{person}{Jingfei Du},
  \bibinfo{person}{Mandar Joshi}, \bibinfo{person}{Danqi Chen},
  \bibinfo{person}{Omer Levy}, \bibinfo{person}{Mike Lewis},
  \bibinfo{person}{Luke Zettlemoyer}, {and} \bibinfo{person}{Veselin
  Stoyanov}.} \bibinfo{year}{2019}\natexlab{}.
\newblock \showarticletitle{Roberta: A robustly optimized bert pretraining
  approach}.
\newblock \bibinfo{journal}{\emph{arXiv preprint arXiv:1907.11692}}
  (\bibinfo{year}{2019}).
\newblock


\bibitem[\protect\citeauthoryear{Mallia, Khattab, Suel, and Tonellotto}{Mallia
  et~al\mbox{.}}{2021}]%
        {mallia2021learning}
\bibfield{author}{\bibinfo{person}{Antonio Mallia}, \bibinfo{person}{Omar
  Khattab}, \bibinfo{person}{Torsten Suel}, {and} \bibinfo{person}{Nicola
  Tonellotto}.} \bibinfo{year}{2021}\natexlab{}.
\newblock \showarticletitle{Learning passage impacts for inverted indexes}. In
  \bibinfo{booktitle}{\emph{Proceedings of the 44th International ACM SIGIR
  Conference on Research and Development in Information Retrieval}}.
  \bibinfo{pages}{1723--1727}.
\newblock


\bibitem[\protect\citeauthoryear{Mustar, Lamprier, and Piwowarski}{Mustar
  et~al\mbox{.}}{2020}]%
        {mustar2020using}
\bibfield{author}{\bibinfo{person}{Agn{\`e}s Mustar}, \bibinfo{person}{Sylvain
  Lamprier}, {and} \bibinfo{person}{Benjamin Piwowarski}.}
  \bibinfo{year}{2020}\natexlab{}.
\newblock \showarticletitle{Using BERT and BART for Query Suggestion.}. In
  \bibinfo{booktitle}{\emph{CIRCLE}}.
\newblock


\bibitem[\protect\citeauthoryear{Nogueira, Lin, and Epistemic}{Nogueira
  et~al\mbox{.}}{2019}]%
        {nogueira2019doc2query}
\bibfield{author}{\bibinfo{person}{Rodrigo Nogueira}, \bibinfo{person}{Jimmy
  Lin}, {and} \bibinfo{person}{AI Epistemic}.} \bibinfo{year}{2019}\natexlab{}.
\newblock \showarticletitle{From doc2query to docTTTTTquery}.
\newblock \bibinfo{journal}{\emph{Online preprint}} (\bibinfo{year}{2019}).
\newblock


\bibitem[\protect\citeauthoryear{Raffel, Shazeer, Roberts, Lee, Narang, Matena,
  Zhou, Li, and Liu}{Raffel et~al\mbox{.}}{2019}]%
        {raffel2019exploring}
\bibfield{author}{\bibinfo{person}{Colin Raffel}, \bibinfo{person}{Noam
  Shazeer}, \bibinfo{person}{Adam Roberts}, \bibinfo{person}{Katherine Lee},
  \bibinfo{person}{Sharan Narang}, \bibinfo{person}{Michael Matena},
  \bibinfo{person}{Yanqi Zhou}, \bibinfo{person}{Wei Li}, {and}
  \bibinfo{person}{Peter~J Liu}.} \bibinfo{year}{2019}\natexlab{}.
\newblock \showarticletitle{Exploring the limits of transfer learning with a
  unified text-to-text transformer}.
\newblock \bibinfo{journal}{\emph{arXiv preprint arXiv:1910.10683}}
  (\bibinfo{year}{2019}).
\newblock


\bibitem[\protect\citeauthoryear{Roitman}{Roitman}{2019}]%
        {roitman2019query}
\bibfield{author}{\bibinfo{person}{Haggai Roitman}.}
  \bibinfo{year}{2019}\natexlab{}.
\newblock \showarticletitle{Query Term Weighting based on Query Performance
  Prediction}.
\newblock \bibinfo{journal}{\emph{arXiv preprint arXiv:1902.10371}}
  (\bibinfo{year}{2019}).
\newblock


\bibitem[\protect\citeauthoryear{Vaswani, Shazeer, Parmar, Uszkoreit, Jones,
  Gomez, Kaiser, and Polosukhin}{Vaswani et~al\mbox{.}}{2017}]%
        {vaswani2017attention}
\bibfield{author}{\bibinfo{person}{Ashish Vaswani}, \bibinfo{person}{Noam
  Shazeer}, \bibinfo{person}{Niki Parmar}, \bibinfo{person}{Jakob Uszkoreit},
  \bibinfo{person}{Llion Jones}, \bibinfo{person}{Aidan~N Gomez},
  \bibinfo{person}{{\L}ukasz Kaiser}, {and} \bibinfo{person}{Illia
  Polosukhin}.} \bibinfo{year}{2017}\natexlab{}.
\newblock \showarticletitle{Attention is all you need}. In
  \bibinfo{booktitle}{\emph{Advances in neural information processing
  systems}}. \bibinfo{pages}{5998--6008}.
\newblock


\bibitem[\protect\citeauthoryear{Wang, Gupta, Hao, Fan, Li, Li, and Guo}{Wang
  et~al\mbox{.}}{2021}]%
        {wang2021contextual}
\bibfield{author}{\bibinfo{person}{Zhuoyi Wang}, \bibinfo{person}{Saurabh
  Gupta}, \bibinfo{person}{Jie Hao}, \bibinfo{person}{Xing Fan},
  \bibinfo{person}{Dingcheng Li}, \bibinfo{person}{Alexander~Hanbo Li}, {and}
  \bibinfo{person}{Chenlei Guo}.} \bibinfo{year}{2021}\natexlab{}.
\newblock \showarticletitle{Contextual Rephrase Detection for Reducing Friction
  in Dialogue Systems}. In \bibinfo{booktitle}{\emph{Proceedings of the 2021
  Conference on Empirical Methods in Natural Language Processing}}.
  \bibinfo{pages}{1899--1905}.
\newblock


\bibitem[\protect\citeauthoryear{Xu and Croft}{Xu and Croft}{2017}]%
        {xu2017quary}
\bibfield{author}{\bibinfo{person}{Jinxi Xu} {and} \bibinfo{person}{W~Bruce
  Croft}.} \bibinfo{year}{2017}\natexlab{}.
\newblock \showarticletitle{Quary expansion using local and global document
  analysis}. In \bibinfo{booktitle}{\emph{Acm sigir forum}},
  Vol.~\bibinfo{volume}{51}. ACM New York, NY, USA, \bibinfo{pages}{168--175}.
\newblock


\bibitem[\protect\citeauthoryear{Zamani, Dehghani, Croft, Learned-Miller, and
  Kamps}{Zamani et~al\mbox{.}}{2018}]%
        {zamani2018neural}
\bibfield{author}{\bibinfo{person}{Hamed Zamani}, \bibinfo{person}{Mostafa
  Dehghani}, \bibinfo{person}{W~Bruce Croft}, \bibinfo{person}{Erik
  Learned-Miller}, {and} \bibinfo{person}{Jaap Kamps}.}
  \bibinfo{year}{2018}\natexlab{}.
\newblock \showarticletitle{From neural re-ranking to neural ranking: Learning
  a sparse representation for inverted indexing}. In
  \bibinfo{booktitle}{\emph{Proceedings of the 27th ACM international
  conference on information and knowledge management}}.
  \bibinfo{pages}{497--506}.
\newblock


\bibitem[\protect\citeauthoryear{Zhang, Chang, Liu, Gollapalli, Li, and
  Xiao}{Zhang et~al\mbox{.}}{2017}]%
        {zhang2017mike}
\bibfield{author}{\bibinfo{person}{Yuxiang Zhang}, \bibinfo{person}{Yaocheng
  Chang}, \bibinfo{person}{Xiaoqing Liu}, \bibinfo{person}{Sujatha~Das
  Gollapalli}, \bibinfo{person}{Xiaoli Li}, {and} \bibinfo{person}{Chunjing
  Xiao}.} \bibinfo{year}{2017}\natexlab{}.
\newblock \showarticletitle{Mike: keyphrase extraction by integrating
  multidimensional information}. In \bibinfo{booktitle}{\emph{Proceedings of
  the 2017 ACM on Conference on Information and Knowledge Management}}.
  \bibinfo{pages}{1349--1358}.
\newblock


\end{thebibliography}

\appendix

\section{Model Hyperparameter}

For the weight predcition model in Figure \ref{fig:model}, Robert-Base\footnote{https://huggingface.co/roberta-base} is used as the pre-trained language model. Multi-head attention \cite{vaswani2017attention} is used as the cross-Attention module, and we set $head = 6$ and $dropout = 0.3$ in this work. A single linear layer is used as the classifier, with a $dropout = 0.5$. Cross-Entropy is used to calculate the loss between each predicted logits and true labels. AdamW is used as the optimizer, with $learning rate = 3e-5$ and $eps = 1e-8$. Training epoch is set as 20, and an early stopping mechanism is applied on the validation step to select the best model.

\section{Evaluation Methods Details}
\begin{figure}[h]
  \centering
  \includegraphics[width=\linewidth]{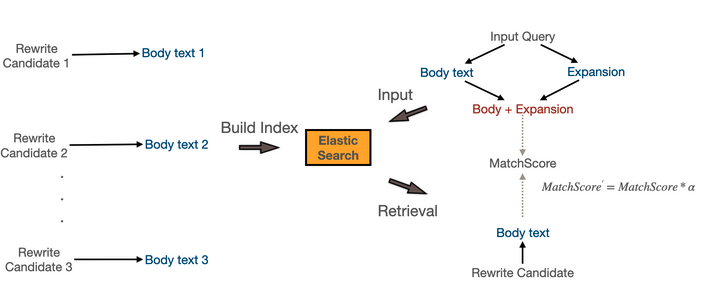}
    \caption{Illustration of using ElasticSearch to measure the model's performance.}
  \label{fig:es}
\end{figure}

Figure \ref{fig:es} demonstrates how we use the ElasticSearch to evaluate the retrieval performance. Each reformulation candidate's body text is input to the ElasticSearch to build the index; given an input query, it will be first concatenated with the expansion to form a new "query + expansion", and this "query + expansion" will be used to retrieval the relevant reformulation candidates. The matching score is calculated based on BM25. After the retrieval, each retrieved reformulation candidate's matching score will be adjusted according to the predicted weight from model \ref{fig:model}: if the candidate contains the entity that is predicted as "2", the candidate's retrieval score will be multiplied by a constant $
\alpha $. We set $\alpha = 1.5$ in this work.

The SimCSE method is a contrastive learning based method. A Roberta-base model is trained using Alexa labelled data with the objective to minimize the distance between positive query-reformulation pairs and maximize the distance between negative pairs. After the training, the model can be used to encode both the query and reformulation candidates to embeddings. We then used the FAISS \footnote{https://github.com/facebookresearch/faiss} to retrieve the most similar reformulation embeddings for each query embedding. Similarly, we adjust each retrieval's score by dividing the original matching distance by $\alpha = 1.2$ if this reformulation contains an important entity. 
\end{document}